\documentclass[aps,prb,reprint,superscriptaddress,showpacs,
amsmath,citeautoscript,flushbottom,floatfix]{revtex4-1}
\usepackage{graphicx}

\newcommand{\vc}[1]{\boldsymbol{#1}}
\newcommand{\ee}{\mathrm{e}}
\newcommand{\SU}{$SU(2)$} 

\setcitestyle{numbers,square}

\allowdisplaybreaks

\begin{document}
\title{Hidden symmetries of the extended Kitaev-Heisenberg model: 
Implications for honeycomb lattice iridates 
${\vc A}_\mathbf{2}$IrO$_\mathbf{3}$}

\author{Ji\v{r}\'{\i} Chaloupka}
\affiliation{Central European Institute of Technology,
Masaryk University, Kotl\'a\v{r}sk\'a 2, 61137 Brno, Czech Republic}

\author{Giniyat Khaliullin}
\affiliation{Max Planck Institute for Solid State Research,
Heisenbergstrasse 1, D-70569 Stuttgart, Germany}

\begin{abstract}
We have explored the hidden symmetries of a generic four-parameter
nearest-neighbor spin model, allowed in honeycomb lattice compounds under
trigonal compression. Our method utilizes a systematic algorithm to identify
all dual transformations of the model that map the Hamiltonian on itself,
changing the parameters and providing exact links between different points in
its parameter space.
We have found the complete set of points of hidden {\SU} symmetry at which
seemingly highly anisotropic model can be mapped back on the Heisenberg model
and inherits therefore its properties such as the presence of gapless
Goldstone modes.
The procedure used to search for the hidden symmetries is quite general and
may be extended to other bond-anisotropic spin models and other lattices, such
as the triangular, kagome, hyper-honeycomb, or harmonic-honeycomb lattices.
We apply our findings to the honeycomb lattice iridates Na$_2$IrO$_3$ and
Li$_2$IrO$_3$, and illustrate how they help to identify plausible values of
the model parameters that are compatible with the available experimental data. 
\end{abstract}

\date{\today}

\pacs{75.10.Jm, %Quantized spin models, including quantum spin frustration
75.25.Dk, %Orbital, charge,and other orders, including coupling of these orders
75.30.Et %Exchange and superexchange interactions
}
\maketitle

\section{Introduction}
%{{{1

When relativistic spin-orbit coupling dominates over the exchange and
orbital-lattice interactions, the orbital moment $\vc{L}$ of an ion remains
unquenched and a total angular momentum $\vc{J}=\vc{S}+\vc{L}$ is formed.
This was known to happen in compounds of late transition metal ions such as 
of cobalt (see, \textit{e.g.}, Ref.~\onlinecite{Buy71}); however, the
``cleanest'' examples of spin-orbit coupled magnets emerged more recently:
these are the iridium oxides Sr$_2$IrO$_4$ and Na$_2$IrO$_3$ with perovskite
and honeycomb lattice structures, correspondingly. 

By construction, magnetic ordering in these systems necessarily involves
interactions between orbital moments $\vc{L}$, in addition to a conventional
Heisenberg exchange among the spin-part of total angular momentum $\vc{J}$
\cite{Ell68}. Since the $L$-moment, hosted by $t_{2g}$ orbital in a crystal,
is only an ``effective'' one \cite{Abr70}, it need not be conserved during the
electron hoppings, thus the $L$-moment exchange interactions are generally not
{\SU} invariant \cite{Kug82}. Moreover, the orbital moments have a ``shape''
and hence the $L$-interactions are anisotropic in real space, too, and thus
strongly frustrated even on simple cubic lattices. Altogether,
this results in nontrivial $L$-Hamiltonians and orderings, including
\textit{e.g.} noncoplanar (multi-Q) states, ``hidden'' Goldstone modes,
\textit{etc.} \cite{Kha02,Kha05}. Via the spin-orbit coupling, these peculiar
features of orbital physics are inherited by the ``pseudospin-$J$''
wavefunctions and interactions
\cite{Kha05,Kha04,Che08,Jac09,Shi09,Cha10,Oka13a,Wit14,Nus13}. In essence, 
frustrated nature and quantum behavior of $t_{2g}$-orbital moments 
\cite{Kha00,Kha01} are transferred to that of low-energy pseudospins $J$. 

Depending on the electron configuration of ions, the ground state pseudospin
may take different values $J=0, 1/2, 1, \ldots$, and a variety of magnetic
Hamiltonians with different symmetries and diverse behavior emerge in each
case, because of different admixture of non-Heisenberg $L$-interactions.
Perhaps the most radical departure from a conventional magnetism is realized
in compounds with apparently ``nonmagnetic'' $J=0$ ions, where a competition 
between spin-orbit and exchange interactions results in a nonmagnetic-magnetic
quantum phase transition \cite{Che09,Kha13,Mee13,noteCoFe}.

The case of pseudospin $J=1/2$ iridates is of special interest. This is
because Sr$_2$IrO$_4$ perovskite was found \cite{Kim09,Kim12,Fuj12} to host
cuprate-like magnetism, and honeycomb lattice iridates $A_2$IrO$_3$
(\mbox{$A=$Na,} Li) have been suggested \cite{Jac09} as a candidate material
where the Kitaev model \cite{Kit06} physics might be realized. Following this
proposal, a subsequent work \cite{Cha10} has introduced the minimal magnetic
Hamiltonian for iridates $A_2$IrO$_3$: the Kitaev-Heisenberg model (KH model)
-- a frustrated spin model with many attractive properties. Most importantly,
its phase diagram contains a finite window of a quantum spin-liquid phase
which emanates from the pure Kitaev point of the model with a known exact
solution \cite{Kit06}. To reflect the later experimental findings in iridates,
such as the zigzag (Na$_2$IrO$_3$ \cite{Liu11,Ye12,Cho12}) and spiral
(Li$_2$IrO$_3$ \cite{Col13}) type magnetic orderings, the initially proposed
model was modified by including longer-range Heisenberg \cite{Kim11,Cho12} or
anisotropic \cite{Reu14} interactions, extending the parameter
range \cite{Cha13,Yu13,Oka13b}, by considering further anisotropic 
terms in the
Hamiltonian \cite{Bha12,Kat14,Rau14a,Rau14b,Yam14,Siz14,Shi14,Kim14a}, or by
including spatial anisotropy of the model parameters \cite{Sel14}. An
alternative picture based on itinerant approach has been also
suggested \cite{Maz12}.

Despite the extensive efforts, no consensus concerning the minimal model for
the honeycomb lattice iridates has thus far been reached. A reliable
microscopic derivation of the exchange interactions is difficult and does not
lead to a conclusive suggestion for the minimal Hamiltonian and its
parameters. On the experimental side, the richest information about the
underlying spin model would be provided by mapping momentum-resolved spin
excitation spectrum. However, due to the lack of large enough monocrystals,
the inelastic neutron scattering (INS) has been performed on powders
only \cite{Cho12}. Another possible probe -- resonant inelastic \mbox{x-ray}
scattering (RIXS) -- suffers from a small resolution at present. While it
could be successfully applied in case of perovskite iridates \cite{Kim12}, here
the limitation comes from the much smaller energy scale of the excitations to
be studied in detail by RIXS; however, the overall strength of magnetic
interactions in Na$_2$IrO$_3$ has been quantified \cite{Gre13,Chu15}.

Nevertheless, the experimental data collected to date puts rather strong
constraints on the possible models. First, the RIXS derived magnetic
energies \cite{Gre13,Chu15} (of the order of 40 meV) are much higher than the
ordering temperature ($\sim 15\:\mathrm{K}$) suggesting strong frustration. 
Second, the
magnetic scattering intensity, measured by RIXS at zero momentum, $\vc Q=0$,
is as strong as elsewhere in the Brillouin zone, which implies a dominance of
anisotropic, non-Heisenberg spin interactions. Third, the recent resonant
x-ray scattering data \cite{Chu15} has revealed nearly ideal $C_3$ symmetry of
the spin correlations in momentum space. 
Moreover, inelastic neutron scattering data \cite{Cho12} have indicated that a
spin gap, if present, would be relatively small (less than $2\:\mathrm{meV}$).
All these observations taken together imply that the dominant
pseudospin interactions in iridates are strongly frustrated, highly
anisotropic in spin space, and yet highly symmetric in real space. By very
construction, all these features are in fact the intrinsic properties of the
KH model and its extended versions.

The KH model, supplemented by other $C_3$ symmetry allowed terms (see below),
is therefore physically sound and plausible. However, there is a problem of
its large parameter space (four parameters even within the nearest-neighbor
model) resulting in complex phase diagrams, which makes the analysis of
experimental data and the extraction of the model parameters a difficult task.
In such cases, clarification of the underlying symmetry properties of the model
is often of a great help. In general, the spin-orbital models in Mott
insulators possess peculiar symmetries \cite{Kha05,Nus13} which are rooted in
the bond-directional nature of orbitals.
In this context, a special four-sublattice rotation \cite{Kha05} within spin
space has proved itself as an extremely useful tool in the case of the
original two-parameter KH model \cite{Cha10,Cha13,Kim14b,Rou12,Bec14,Li14}. 
It maps the
Hamiltonian on itself but changes the Hamiltonian parameters, connecting
thereby different points in the parameter space. Being an exact
transformation, it transfers the complete knowledge about some point in the
phase diagram, including the groundstate, excitation spectrum, response
functions \textit{etc.}, to its partner. Based solely on this self-duality of
the model, the entire phase diagram could be sketched and the deep relations
between the phases understood. In addition, it also reveals points of hidden
{\SU} symmetry, where the system is exactly equivalent to a Heisenberg model
for the rotated spins. Given its usefulness, it is highly desirable to find
and analyze similar transformations for the extended versions of the KH model. 

In this paper, we introduce a systematic method to derive dual transformations
of bond-anisotropic spin Hamiltonians and demonstrate its results and their
physical implications in the case of honeycomb iridates adopting the full
nearest-neighbor model \cite{Kat14,Rau14a,Yam14}. We find all the hidden
{\SU}-symmetry points of the model, the most peculiar one being characterized
by a ``vortex''-like pattern with a six-site unit cell, and demonstrate how
the characteristics of the hidden Heisenberg magnet manifest themselves in the
anisotropic situations. By identifying the {\SU} points we characterize all
the possible gapless Goldstone modes that may be encountered within the model.
This is relevant in the context of real materials as the spin gap was found to
be well below $2\:\mathrm{meV}$ \cite{Cho12,Col13}, suggesting a connection to
some of the {\SU} points. Finally, using a self-duality of the model, we will
provide a link between our fits of the earlier Na$_2$IrO$_3$ data \cite{Cha13}
and the recent experimental observation of the ordered moment direction
\cite{Chu15}. We argue that this observation provides a direct access to the
strength of the additional terms ``extending'' the KH model, and quantify the
spin easy axis direction in terms of this ``departure'' from the pure KH
model. This allows us to suggest plausible values of the model parameters that
are compatible with the current data. While we focus here on the case of a
honeycomb lattice as realized in Na$_2$IrO$_3$ and more recently in RuCl$_3$
\cite{Plu14}, the method is general and expected to produce interesting
results also in the context of the new structural families of iridates --
recently synthesized hyper-honeycomb \cite{Tak14,Bif14b} and
harmonic-honeycomb lattices \cite{Mod14,Bif14a}, or the theoretically proposed
hyperoctagon lattice \cite{Her14}.

The paper is organized as follows. Sec. II introduces the Hamiltonian and
discusses its parameters. Sections III and IV introduce the method, derive 
and discuss the main results of the paper -- the hidden symmetries of the 
model. Sec. V and Appendix B discuss the implications of the results for 
honeycomb iridates.
%}}}1

% ==================================================

\section{Extended Kitaev-Heisenberg model}\label{sec:model}
%{{{1

%-figure 1-----------------------------------------------------------------
\begin{figure}[tb]
\begin{center}
\includegraphics[scale=1.00]{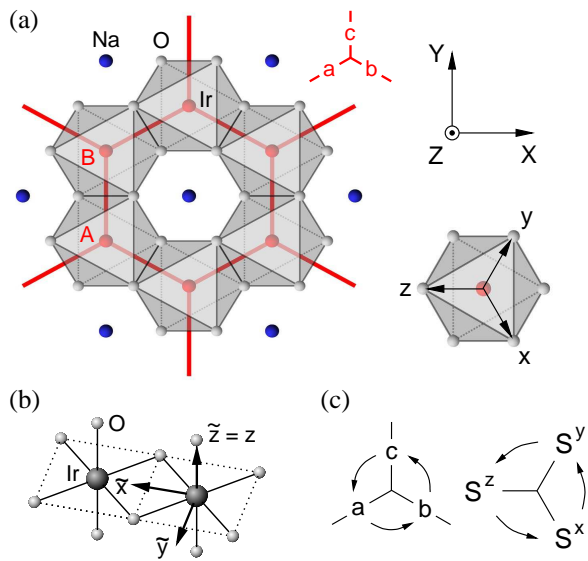}
\caption{(Color online)
(a)~Top view of the honeycomb NaIr$_2$O$_6$ plane, the definition of global
$X,Y,Z$ axes, and the $xyz$ reference frame for the spin components.
The $X$- and $Y$-directions coincide with the crystallographic
${\bf a}$- and ${\bf b}$-axes. The three bond directions of the honeycomb
lattice are labeled as $a$, $b$, and $c$, its two sublattices are 
labeled by $A$ and $B$.
(b)~Two edge-shared IrO$_6$ octahedra of a \mbox{$c$-bond} and the 
definition of the local spin axes $\tilde{x}$, $\tilde{y}$, $\tilde{z}$ 
(used in Eq.~\ref{eq:HXYZ}). 
(c)~Simultaneous cyclic permutation of the Ir-Ir bond directions 
$a$, $b$, $c$ and the spin components $x$, $y$, $z$ when applying 
a $C_3$ rotation to the model.
}
\label{fig:schem}
\end{center}
\end{figure}
%--------------------------------------------------------------------------

We start by specifying the model Hamiltonian including all symmetry-allowed
spin interactions on nearest-neighbor bonds. An ideal, undistorted, structure
of the honeycomb NaIr$_2$O$_6$ plane is shown in Fig.~\ref{fig:schem}(a). We
will utilize its rotational $C_3$ symmetry and the three sets of parallel
mirror planes containing the shared edges of the IrO$_6$ octahedra and cutting
the Ir-Ir bonds into halves. The $C_3$ symmetry links the interactions for
different bond directions while the presence of the mirror planes restricts
the possible interactions for a given bond direction. A trigonal distortion
(compression or elongation along the \mbox{$Z$-axis}) fully preserves these
symmetries so that our Hamiltonian applies in that case as well. Furthermore,
recent experiments \cite{Chu15} indicate a nearly ideal $C_3$ symmetry of the
spin properties and hence suggest that additional terms, possibly induced by a
monoclinic distortion present in Na$_2$IrO$_3$, can be neglected.
Physically, this observation implies the robustness of the pseudospin
wavefunctions against weak monoclinic distortions.

The bond Hamiltonian is most compactly expressed in a local, bond-dependent,
$\tilde{x} \tilde{y} \tilde{z}$ reference frame for spins, presented in
Fig.~\ref{fig:schem}(b) for a $c$-bond. Due to the mirror symmetry, the
in-bond $S^{\tilde{x}}$ component is forbidden to interact with the
$S^{\tilde{y}}$ and $S^{\tilde{z}}$ components \cite{Kat14}. Following the 
notation of Ref.~\onlinecite{Kat14} we arrange the allowed terms into the form
\begin{multline}\label{eq:HXYZ}
\mathcal{H}_{\langle ij\rangle \,\parallel\, c} =J\, {\vc S}_i \cdot {\vc S}_j 
+ K\, S_i^{\tilde{z}} S_j^{\tilde{z}} \\
+ D\, (S_i^{\tilde{x}} S_j^{\tilde{x}} - S_i^{\tilde{y}} S_j^{\tilde{y}}) 
+ C\, (S_i^{\tilde{y}} S_j^{\tilde{z}} + S_i^{\tilde{z}} S_j^{\tilde{y}}) \;.
\end{multline}
This four-parameter Hamiltonian extends the KH model ($J$- and $K$-terms) by
the $D$-term bringing further anisotropy among the diagonal components of the
interaction, and the $C$-term determining the only symmetry-allowed
non-diagonal element in the exchange interaction tensor. Parameter $C$
would vanish for an isolated pair of undistorted octahedra; it becomes finite
due to a trigonal distortion and/or due to the extended nature of orbitals 
in a crystal (``recognizing'' the fact that the octahedra are canted relative 
to the crystal axis $Z$). 

To capture the $C_3$
symmetry, it is convenient to switch to cubic axes $xyz$, introduced in
Fig.~\ref{fig:schem}(a) and pointing from an Ir ion to neighboring O ion
positions in an ideal structure. The $c$-bond Hamiltonian in the cubic
reference frame, as derived in Ref.~\onlinecite{Rau14a}, reads then as: 
\begin{multline}\label{eq:Hcubic}
\mathcal{H}_{\langle ij\rangle \,\parallel\, c} = 
J\, {\vc S}_i \cdot {\vc S}_j + K\, S_i^z S_j^z \\
\!+\!\Gamma(S_i^x S_j^y \!+\! S_i^y S_j^x)
\!+\!\Gamma'(S_i^x S_j^z \!+\! S_i^z S_j^x \!+\! S_i^y S_j^z \!+\! S_i^z S_j^y),
\end{multline}
with the correspondence $\Gamma=-D$ and $\Gamma'=\frac{1}{\sqrt{2}}C$, often
used below. For the other bond directions, the
Hamiltonian is obtained by a cyclic permutation [see Fig.~\ref{fig:schem}(c)],
resulting in one-to-one correspondence between the three types of bonds and
interactions, as required by $C_3$ symmetry. Physically, each type of bonds
favors its own distinct ``orbital setup'' to optimize the hopping energy, and
this is fingerprinted in pseudospin interactions via spin-orbit coupling. 
For completeness, the Appendix A shows the Hamiltonian in the global axes 
$XYZ$; it has certain advantages moving the bond dependence from the operator
forms to the coupling constants. 

Few comments are in order concerning the model parameters. In general,
calculation of exchange integrals in transition metal compounds with
$90^\circ$ \mbox{$d$-$p$-$d$} bonding geometry is an intricate task, because
more hopping pathways are allowed as compared to a simpler case of $180^\circ$
\mbox{$d$-$p$-$d$} bonding in perovskites (where theory \cite{Jac09} has
correctly predicted the strength of dominant exchange constants). For
instance, $t_{2g}$ orbitals may also overlap directly, in addition to oxygen
mediated hoppings; there is a large overlap between orbitals of $t_{2g}$ and
$e_g$ symmetries (forbidden in perovskites), \textit{etc.}, resulting in a
number of competing ferromagnetic and antiferromagnetic contributions which
are difficult to evaluate, in particular in compounds with small Mott and/or
charge-transfer excitation gaps. The uncertainties in interaction parameters
$U$ and $J_H$ further affect the theoretical estimates.
 
Initial consideration \cite{Kha05} of the pseudospin one-half exchange 
interactions in $90^\circ$-bonding geometry resulted in $K=-2J$ 
(hitting a ``hidden'' {\SU} point by chance) in the cubic limit; 
later work \cite{Che08,Jac09,Cha10} using different approximations has 
changed this estimate both in terms of the signs and values of $J$ and $K$, 
illustrating the difficulties described above. It was also found that the 
non-diagonal element $\Gamma$ allowed in cubic symmetry may take
sizable values \cite{Rau14a,Kat14,Yam14,Siz14}. 
Further, $\Gamma'$ is expected to
become as large as the other parameters if trigonal splitting $\Delta$ of the
$t_{2g}$ orbital level, caused by a compression along \mbox{$Z$-axis}, becomes
comparable to spin-orbit coupling $\lambda$; also, the trigonal field
suppresses the parameter $K$. These trends are easy to understand: large
trigonal field suppresses the in-plane components of orbital moment $L_{X}$
and $L_{Y}$, leaving the axial $L_{Z}$ component the only unquenched one; thus
the pseudospin one-half Hamiltonian, written most conveniently in global axes
in this limit, may not contain anything but $XX+YY$ and $ZZ$ type terms: 
$J_{XY} (S_i^X S_j^X \!+\! S_i^YS_j^Y) + J_Z S_i^ZS_j^Z$,
identical for all bonds. This is what has indeed been found by explicit
calculations \cite{Kha05,Bha12} in the limit of $\Delta\gg\lambda$. This implies
$K=0$ and $\Gamma'=\Gamma$ in this limit (see also Appendix A), 
while $J_{XY}=(J-\Gamma)$ and $J_Z=(J+2\Gamma)$ may take any values depending
on the microscopic details.
Although this limit is not very realistic for Ir$^{4+}$ ion with large
spin-orbit constant $\lambda\sim 0.4\:\mathrm{eV}$ \cite{Abr70,Fig00}, we may
expect sizable values of both $\Gamma$ and $\Gamma'$ in Na$_2$IrO$_3$ where
$\Delta$ seems to exceed $0.1\:\mathrm{eV}$ \cite{Gre13a,noteDelta}.
The role of $\Gamma$ and $\Gamma'$ terms should further increase in other 
compounds based on pseudospin one-half Co$^{4+}$, Ru$^{3+}$, and Rh$^{4+}$ 
ions with smaller $\lambda$.

In general, the high-energy behavior of spins and orbitals in transition metal
compounds is well captured by the Kugel-Khomskii models \cite{Kug82} and their 
descendants \cite{Kha05}. However, the low-energy physics and ultimate magnetic 
``fixed-point'' are heavily influenced by many ``unpleasant'' details
originating from orbital-lattice coupling and distortions, unavoidable 
in real materials. In perovskites, the Kugel-Khomskii energy scale is given by 
$4t^2/U$ independent on spin-orbit coupling; however, this leading 
term drops out for pseudospins one-half in the edge-shared, $90^\circ$-bonding
geometry \cite{Kha05,Jac09}, so the ``high-energy'' scale is set up by 
the subleading terms. In iridates, the hope \cite{Jac09} is that the 
Kitaev-type coupling is the leading one among these subleading terms.
Since this coupling is itself a correction
to $4t^2/U$, this expectation may or may not hold in reality.

To summarize up to now: in real materials even with an ideal $C_3$ symmetry,
all the four exchange parameters may play a significant role. This motivates
us to regard the Hamiltonian \mbox{(\ref{eq:HXYZ},\ref{eq:Hcubic})} as an
effective model with arbitrary parameters, and look for some general symmetry
arguments that may help to identify plausible parameter windows in the
analysis of experimental data.

%}}}1

% ==================================================

\section{Systematic construction of dual transformations}
%{{{1

Having fixed the model Hamiltonian, we are ready to explore its dual
transformations. By a dual transformation we mean a prescription for
site-dependent rotations in the spin space, $\vc S'_i=\mathcal{R}_i \vc S_i$,
which transforms a spin Hamiltonian $\mathcal{H}(\vc S)$ into a formally new
Hamiltonian $\mathcal{H}'(\vc S')$. We are interested in self-dual
transformations of $\mathcal{H}$ that map the model onto itself, preserving
its all symmetry properties. That is, the rotated partner $\mathcal{H}'$ 
({\it i})~has the same four terms albeit with different parameters $J'K'D'C'$, 
and ({\it ii})~it respects the $C_3$ rotation rules encoded in
Fig.~\ref{fig:schem}(c), hence preserving the original distribution of the
three types of bond-dependent interactions on a lattice.

Starting with the $JKDC$ Hamiltonian expressed as
$\mathcal{H}(\vc S)=\sum_{\langle ij\rangle} {\vc S}_i^T H_{ij} {\vc S}_j$ 
where $H_{ij}$ are $3\times 3$ matrices, we obtain
$\mathcal{H}(\vc S)=\mathcal{H}'(\vc S')=\sum_{\langle ij\rangle} 
{{\vc S}'_i}^T H'_{ij} {\vc S}'_j$ with
$H'_{ij}=\mathcal{R}_i H_{ij} \mathcal{R}_j^T$. For a self-dual
transformation, the matrices $H'_{ij}$ are identical to $H_{ij}$, but the 
parameters $JKDC$ are replaced by $J'K'D'C'$, and the one-to-one correspondence 
between the bond directions and interactions remains intact.
These two points in the parameter space are linked by the transformation and
knowing the solution at one of the points, we may ``rotate'' it to the other
one.

In this section we give an algorithm to find the self-dual transformations for
the extended KH model that map it onto itself. We have found a single
self-dual transformation $JKDC\leftrightarrow J'K'D'C'$ operating in full
parameter space of the model; we will show it shortly below and return to it
later when discussing experimental data. 

However, studying the hidden symmetries of the model, we have identified a
number of {\it restricted} self-dual transformations that operate only in some
regions of the parameter space, where constants $J,K,D,C$ are all finite but
obey certain relations, or some of them are simply zero. Our primary interest
is in the special class of such transformations of the type
$J_0\leftrightarrow JKDC$, which convert the Heisenberg model into the full
$JKDC$ model and vice versa. These transformations, to be discussed in the
next section, reveal points of hidden {\SU} symmetry -- by inverting the
transformation the anisotropic model with the parameters $JKDC$ can be exactly
mapped back to the Heisenberg model with the exchange constant $J_0$.

\subsection{Algorithm}

A systematic search for the dual transformations seems to be an intricate
task. Fortunately, it can be easily performed by computer on a finite cluster
of the lattice using the following simple algorithm. We give it specifically
for the case of a self-dual transformation: 

{\bf A})~As a first step, we choose two rotation matrices $\mathcal{R}_i$,
$\mathcal{R}_j$ on a selected bond $\langle ij\rangle$. They have to preserve
the $JKDC$-form given by \eqref{eq:HXYZ}, which leaves us with only a few
choices, each having only one free angular parameter. 

{\bf B})~Next, we randomly choose nonzero values of the initial parameters
$JKDC$ and use the relation $H'_{ij}=\mathcal{R}_i H_{ij} \mathcal{R}_j^T$
together with the $C_3$ symmetry to determine the new Hamiltonian matrices for
the three bond directions. 

{\bf C})~Knowing all the bond Hamiltonians, we may now determine further
rotation matrices by utilizing relations of the type
$\mathcal{R}_j=(H'_{ij})^{-1}\mathcal{R}_i H_{ij}$ and proceeding
neighbor-by-neighbor. To fully determine the rotation matrices,
about two thirds of the bonds need to be used. 

{\bf D})~The bonds of the remaining third are used to check consistency, the
Hamiltonian matrix determined by using the rotation matrices belonging to the
bond has to be identical to that determined in step~{\bf B}. If the total 
difference on all the remaining bonds equals zero, we have just constructed 
a self-dual transformation. By scanning through the entire interval of the 
free parameter introduced in step~{\bf A}, we find all the self-dual 
transformations.

The above procedure may be easily adapted to find the dual transformations
such as $J_0\leftrightarrow JKDC$. In this case, in step~{\bf A} of the
algorithm, we use the symmetry of the Heisenberg model and choose
$\mathcal{R}_i$ as an identity matrix. The choice of the second matrix
$\mathcal{R}_j$ is restricted by the requirement that 
$H'_{ij}=\mathcal{R}_i H_{ij} \mathcal{R}_j^T=J\mathcal{R}_j^T$ 
is of the $JKDC$ form. 

By inspecting the rotation matrices of the cluster, we can identify the
particular unit cell of the transformation. Note, that even if our cluster is
smaller than this unit cell, we do not miss the corresponding transformation,
so that the method is completely systematic \cite{noteKpts}.

\subsection{Self-duality of the extended Kitaev-Heisenberg model}

The systematic procedure described above has identified only a single
self-dual transformation $JKDC\leftrightarrow J'K'D'C'$. This is not
surprising given the complexity of the model. The corresponding parameter
transformation may be written in a matrix form
\begin{equation}\label{eq:fulldual}
\begin{pmatrix}
J' \\ K' \\ D' \\ C' 
\end{pmatrix}
=
\begin{pmatrix}
1 & +\frac49 & +\frac49 & +\frac{2\sqrt{2}}9 \\
0 & -\frac13 & -\frac43 & -\frac{2\sqrt{2}}3 \\
0 & -\frac49 & +\frac59 & -\frac{2\sqrt{2}}9 \\
0 & -\frac{2\sqrt{2}}9 & -\frac{2\sqrt{2}}9 & +\frac79
\end{pmatrix}
\begin{pmatrix}
J \\ K \\ D \\ C 
\end{pmatrix}\;.
\end{equation}
For convenience, we also give the transformation of the parameters
$JK\Gamma\Gamma'$ entering the Hamiltonian \eqref{eq:Hcubic}:
\begin{equation}\label{eq:fulldualG}
\begin{pmatrix}
J \\ K \\ \Gamma \\ \Gamma' 
\end{pmatrix}'
=
\begin{pmatrix}
1 & +\frac49 & -\frac49 & +\frac49                 \\
0 & -\frac13 & +\frac43 & -\frac43\rule{0pt}{11pt} \\
0 & +\frac49 & +\frac59 & +\frac49\rule{0pt}{11pt} \\
0 & -\frac29 & +\frac29 & +\frac79\rule{0pt}{11pt}
\end{pmatrix}
\begin{pmatrix}
J \\ K \\ \Gamma \\ \Gamma'
\end{pmatrix}\;.
\end{equation}
In terms of the spins, the transformation, labeled for future reference as
$\mathcal{T}_1$, is simply a global \mbox{$\pi$-rotation} about the
\mbox{$Z$-axis} defined in Fig.~\ref{fig:schem}(a). The individual $S^X$,
$S^Y$, and $S^Z$ components transform according to
\begin{equation}
\mathcal{T}_1:\quad (X',Y',Z')=(-X,-Y,Z)
\end{equation}
at every site. By applying the transformation twice, we get an identity and
the matrices in \mbox{(\ref{eq:fulldual},\ref{eq:fulldualG})} are thus 
self-inverse. Despite its apparent
triviality, this transformation will play an essential role when discussing
the real materials, see Sec.~\ref{sec:mater} below.

%}}}1

% ==================================================

\section{Points of hidden SU(2) symmetry}
%{{{1

In this paragraph we find and characterize all the points of hidden {\SU}
symmetry present in the extended KH model. At these special points in the
parameter space, the anisotropic model can be mapped back to a Heisenberg
ferromagnet or antiferromagnet. The {\SU} points of the original KH model have
been identified \cite{Cha10} by virtue of the four-sublattice transformation
introduced in Ref.~\onlinecite{Kha05}. The corresponding ordering patterns on
the honeycomb lattice are of stripy and zigzag type. A similar symmetry
analysis of the KH model was performed for other relevant
lattices \cite{Kim14b}.

The extended KH model of course inherits the {\SU} points of the KH model and
contains several new ones in addition. They are identified by dual
transformations of the type $J_0\leftrightarrow JKDC$ which is
less general than $JKDC\leftrightarrow J'K'D'C'$.
Because of this, we obtain a
relatively rich set of dual transformations characterized by two-, four-, and
six-sublattice structure of the rotations. In terms of parameters, all the
non-trivial {\SU} points are listed in Table~\ref{tab:SU2}. We now proceed 
with the detailed description of the corresponding transformations. 

%-table 1------------------------------------------------------------------
\begin{table}[tb]
\begin{ruledtabular}
\begin{tabular}{lrrrr}
 & $J/J_0$ & $K/J_0$ & $(\Gamma\!\equiv\!-D)/J_0$ & 
$(\Gamma'\!\equiv\!\frac{1}{\sqrt{2}}C)/J_0$ \\
\colrule
$\mathcal{T}_2$              & $-1/3$ & $0$ & $2/3$ & $2/3$ \\
$\mathcal{T}_4$              & $-1$ & $2$ & $0$ & $0$ \\
$\mathcal{T}_6$              & $0$ & $-1$ & $-1$ & $0$ \\
$\mathcal{T}_1\mathcal{T}_4$ & $-1/9$ & $-2/3$ & $8/9$ & $-4/9$ \\
$\mathcal{T}_2\mathcal{T}_6$ & $-2/3$ & $1$ & $1/3$ & $-2/3$ 
\end{tabular}
\end{ruledtabular}
\caption{Parameter values for the {\SU} points in units of the exchange
constant $J_0$ of the hidden Heisenberg model.}
\label{tab:SU2}
\end{table}
%--------------------------------------------------------------------------

\subsection{Summary of the SU(2) points and the corresponding rotations 
on the sublattices}
%{{{2

\begin{figure}[t!b]
\begin{center}
\includegraphics[scale=1.00]{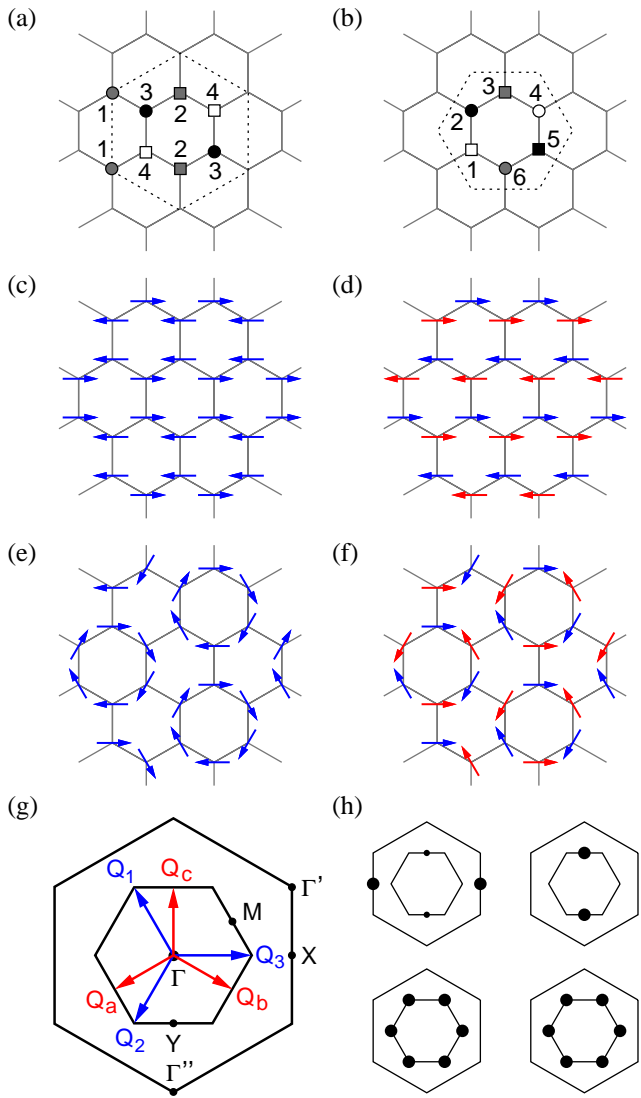}
\caption{(Color online)
(a,b)~Unit cells for the four- and six-sublattice transformations.
(c,d)~Stripy and zigzag patterns related to the FM and AF order of a hidden
Heisenberg magnet via the four-sublattice transformation $\mathcal{T}_4$. The
spins take the \mbox{$z$-axis} direction.
(e,f)~``Vortex''-like patterns generated by the six-sublattice transformation
$\mathcal{T}_6$. The spins are lying in the lattice plane in the case
presented. The colors of the arrows in panels (d) and (f) indicate the
sublattices of the hidden AF order.
(g)~Brillouin zones of the honeycomb (inner hexagon) and the completed
triangular lattice (outer hexagon). The characteristic vectors $\vc Q_{a,b,c}$
of the four-sublattice transformation and $\vc Q_{1,2,3}$ of the
six-sublattice transformation are shown in red and blue, respectively.
(h)~Bragg spots of the patterns in panels (c-f). The dot size is proportional
to $|S_{\vc Q}|$.
}
\label{fig:pattern}
\end{center}
\end{figure}

We first give a summary of the transformations as represented by rotations in
the real space. Each of them generates an infinite number of orderings, since
the ordered moment direction in the underlying Heisenberg model can be chosen
arbitrarily. Figure~\ref{fig:pattern} shows a few important examples.

%-figure 2-----------------------------------------------------------------
%--------------------------------------------------------------------------

The simplest transformation $\mathcal{T}_2$ is $\pi$-rotation about
\mbox{$Z$-axis} at one of the two sublattices of the honeycomb lattice:
\begin{align}
\mathcal{T}_2:&& \quad
   (X',Y',Z')&=( X, Y,Z) & \text{(sublattice A)}\;, \notag \\
&& (X',Y',Z')&=(-X,-Y,Z) & \text{(sublattice B)}\;. 
\end{align}
Its physical relevance is small due to the dominance of 
$\Gamma'(\equiv \frac1{\sqrt 2}C)$ and the complete 
absence of $K$ (corresponding to the case of strong trigonal field 
splitting, as explained above). As a curiosity, if we choose the spins to 
lie in the honeycomb plane, $\mathcal{T}_2$ converts the FM pattern to 
AF and vice versa. We may thus have an AF/FM ordered pattern but the hidden
nature revealing itself \textit{e.g.} in the spin dynamics is that of
Heisenberg FM/AF, respectively.

The next transformation $\mathcal{T}_4$ has a four sublattice structure
depicted in Fig.~\ref{fig:pattern}(a) with $\pi$-rotations about cubic $x$,
$y$, and $z$ axes applied at sublattices $1$, $2$, and $3$, respectively, and
no rotation involved at sublattice $4$. Written explicitly:
\begin{align}
\mathcal{T}_4:&& \quad
   (x',y',z')&=(x,-y,-z) & \text{(sublattice 1)}\;, \notag \\
&& (x',y',z')&=(-x,y,-z) & \text{(sublattice 2)}\;, \notag \\
&& (x',y',z')&=(-x,-y,z) & \text{(sublattice 3)}\;, \notag \\
&& (x',y',z')&=(x, y, z) & \text{(sublattice 4)}\;. 
\end{align}
This transformation, introduced earlier in Ref.~\onlinecite{Kha05}, is a
self-dual transformation of the original two-parameter KH model and has been
already heavily used in this context. Applying the transformation to an
ordered Heisenberg FM/AF with the moments pointing along the \mbox{$z$-axis}, 
we get the stripy/zigzag order shown in Fig.~\ref{fig:pattern}(c,d).

Perhaps the most surprising {\SU} point of the model is linked to the
six-sublattice transformation $\mathcal{T}_6$. Its rotations are most
conveniently described in the cubic coordinates. On the lattice sites $1$,
$3$, and $5$ [see Fig.~\ref{fig:pattern}(b)] they correspond to cyclic 
permutations among the spin components.
On the lattice sites $2$, $4$, and $6$ the rotations correspond to anti-cyclic
permutations which have to be followed by a spin-inversion. Altogether the
transformation can be written as
\begin{align}
\mathcal{T}_6:&& \quad
   (x',y',z')&=( x, y, z) & \text{(sublattice 1)}\;, \notag \\
&& (x',y',z')&=(-y,-x,-z) & \text{(sublattice 2)}\;, \notag \\
&& (x',y',z')&=( y, z, x) & \text{(sublattice 3)}\;, \notag \\
&& (x',y',z')&=(-x,-z,-y) & \text{(sublattice 4)}\;, \notag \\
&& (x',y',z')&=( z, x, y) & \text{(sublattice 5)}\;, \notag \\
&& (x',y',z')&=(-z,-y,-x) & \text{(sublattice 6)}\;.
\end{align}
It is easy to see, that for $K=\Gamma(\equiv -D)$ and $J=\Gamma'=0$, these
rotations lead to the isotropic Heisenberg Hamiltonian. As an example, we
consider the \mbox{$c$-bond} $1$-$2$ of Fig.~\ref{fig:pattern}(b). By
exchanging $x$ and $y$ at site $2$, the non-diagonal $\Gamma$-term in
\eqref{eq:Hcubic} becomes diagonal and the inversion ensures its proper sign.
Sample patterns generated by $\mathcal{T}_6$ and showing a ``vortex''-like
structure are presented in Fig.~\ref{fig:pattern}(e,f). The peculiarity of
the {\SU} points is now best demonstrated: the Hamiltonian is completely
anisotropic containing $K$ and $\Gamma(\equiv -D)$ terms only, the ordered 
spins form a very unusual pattern, yet the hidden nature of the system is 
exactly that of the Heisenberg FM or AF, including \textit{e.g.} the presence 
of gapless Goldstone modes.
 
Apart from revealing a hidden {\SU} point of the present model, 
the $\mathcal{T}_6$ transformation has a remarkable property 
that deserves a special attention. Namely, 
applying $\mathcal{T}_6$ to the Kitaev Hamiltonian, we notice that it 
re-distributes three types of Ising-interactions on a honeycomb lattice such 
that at each hexagon a Kekul\'{e} type pattern is formed \cite{Kam10}. 
We thus arrive at the so-called Kekul\'{e}-Kitaev model \cite{Qui15}. 
In other words, the Kitaev and Kekul\'{e}-Kitaev models are exact dual 
partners linked via the $\mathcal{T}_6$ transformation. This observation 
should be helpful in studying both models, in particular of their extended 
versions including a Heisenberg term \cite{Qui15,noteT6Kek}.

Two more transformations providing {\SU} points are obtained as the
combinations $\mathcal{T}_1\mathcal{T}_4$ and $\mathcal{T}_2\mathcal{T}_6$.
They share the sublattice structure with $\mathcal{T}_4$ and $\mathcal{T}_6$,
respectively. Adopting the extended KH model, the former one is probably the
{\SU} point closest to the real situation in Na$_2$IrO$_3$ as will be
discussed in Sec.~\ref{sec:mater}.

%}}}2

\subsection{Implications for the phase diagram}\label{sec:PD}
%{{{2

After examining the nature of the individual {\SU} points, we want to
visualize now their positions in the parameter space, get a sketch of the
phase diagram, and infer the relations between the individual phases. The
result can be compared with the published phase diagrams of
Refs.~\onlinecite{Rau14a} and \onlinecite{Rau14b}, obtained by classical
analysis and partly complemented by exact diagonalization. For this reason,
we adopt the representation of the parameter space introduced in
Ref.~\onlinecite{Rau14a}. The overall energy scale irrelevant for the phase
diagram is removed and $J$, $K$, $\Gamma$ are parametrized using ``spherical''
angles $\theta$ and $\phi$ via $J=\sin\theta\cos\phi$, $K=\sin\theta\sin\phi$,
and $\Gamma=-D=\pm\cos\theta$, keeping $\Gamma' (\equiv\frac{1}{\sqrt{2}}C)$
as a separate parameter of the phase portrait.

%-figure 3-----------------------------------------------------------------
\begin{figure}[tb]
\begin{center}
\includegraphics[scale=1.00]{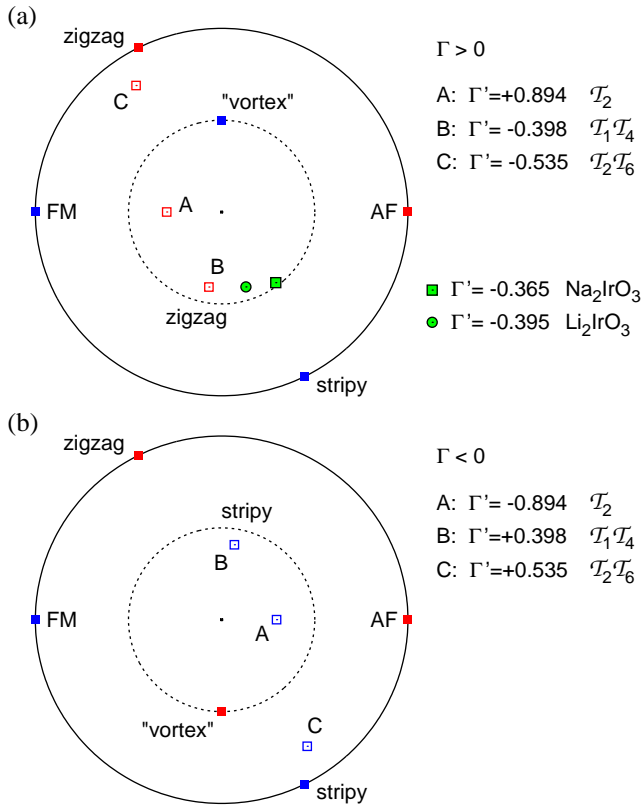}
\caption{(Color online)
(a)~Depiction of the {\SU} points using the parametrization of
Ref.~\onlinecite{Rau14a}, $J=\sin\theta\cos\phi$,
$K=\sin\theta\sin\phi$, $\Gamma=\cos\theta$. The distance from the center of
the circle corresponds to $\theta$ going from $0$ (center) through $\pi/4$
(dashed circle) to $\pi/2$ (solid circle). The polar angle is $\phi$. Filled
squares show the {\SU} points with $\Gamma'=0$, open squares those with
nonzero $\Gamma'$ values given on the right along with the transformation
label. The color of the points indicates their hidden FM (blue) or AF (red)
nature. The green square (circle) shows the parameter values specified in
Sec.~\ref{sec:mater} when discussing Na$_2$IrO$_3$ (Li$_2$IrO$_3$). 
(b)~The same as in panel (a) but with $\Gamma=-\cos\theta$.
}
\label{fig:PD}
\end{center}
\end{figure}
%--------------------------------------------------------------------------

Shown in Fig.~\ref{fig:PD} is the complete set of {\SU} points of the extended
KH model. The outer rings correspond to the original KH model and contain the
trivial {\SU} points and the two well-known $\mathcal{T}_4$ hidden {\SU}
points of the KH model characterized by a stripy and zigzag pattern. Still
within the $JK\Gamma$ plane is the ``vortex'' $\mathcal{T}_6$ point associated 
with a ``vortex''-like pattern. The corresponding phases determined by these
{\SU} points can be observed in Figs.~2 and 3 of Ref.~\onlinecite{Rau14a}, with 
the $\mathcal{T}_6$ point lying in their $120^\circ$ phase.

Three more {\SU} points $A$, $B$, and $C$ characterized by a nonzero value of
$\Gamma'$ are shown as projected onto the $JK\Gamma$ plane. For the $\Gamma>0$
case presented in Fig.~\ref{fig:PD}(a), they are of AF character, one of them
appears for positive (point $A$) and two for negative (points $B$ and $C$)
values of $\Gamma'$. The point $A$ (given by $\mathcal{T}_2$) of hidden AF
nature can possess FM pattern as discussed in the previous paragraph. The
region between the true FM Heisenberg point and the point $A$ in the phase
diagram obtained classically is therefore filled by the FM phase extending as
$\Gamma'$ increases [see panels (c) and (e) of Fig.~2 of
Ref.~\onlinecite{Rau14b}]. However, the (hidden) nature of this phase changes
from FM to AF which should manifest itself \textit{e.g.} on the character of
the magnon dispersion. Similarly, the presence of the points $B$
($\mathcal{T}_1\mathcal{T}_4$) and $C$ ($\mathcal{T}_2\mathcal{T}_6$) of
zigzag and ``vortex'' character, respectively, explains the enlarged region of
the corresponding phases in the classical phase diagram for $\Gamma'<0$ [see
the panels (a) and (d) of Fig.~2 of Ref.~\onlinecite{Rau14b}]. We also observe
an intimate relation between the zigzag phase emanating from the $B$
($\mathcal{T}_1\mathcal{T}_4$) point and that connected to the zigzag {\SU}
point of the original KH model (given by $\mathcal{T}_4$). Due to the
additional $\mathcal{T}_1$ rotation, their ordered moment directions are
related by $\pi$-rotation about the global \mbox{$Z$-axis}. This point will be
further discussed in Sec.~\ref{sec:mater}.  Finally, similar conclusion as for
the $\Gamma>0$ case presented in Fig.~\ref{fig:PD}(a) can be drawn for the
$\Gamma<0$ case shown in Fig.~\ref{fig:PD}(b). The {\SU} points are related by
inversion with respect to the center of the circle and the opposite FM/AF
nature.

In summary, we have illustrated that the gross features of the phase diagram
of the extended, four-parameter KH model can be deduced solely by inspecting 
the nature of the points of hidden {\SU} symmetry and their location 
in the parameter space.

%}}}2

\subsection{Spin excitation spectra}
%{{{2

We proceed further by inspecting the spin excitation spectra at the {\SU}
points associated with $\mathcal{T}_4$ and $\mathcal{T}_6$ transformations,
and see how they are related to those of the simple Heisenberg model. To this
end, the dual transformations have to be expressed in Fourier space and
relations between the Fourier components $\vc S_{\vc q}$ of the dual partners
have to be established. The situation is somewhat complicated by the
two-sublattice structure of the honeycomb lattice, requiring us to introduce
an additional index [see the labels $A$ and $B$ in Fig.~\ref{fig:schem}(a)
for the convention used below].

In both cases, it is convenient to use the cubic axes $xyz$. The
four-sublattice transformation has three characteristic vectors $\vc
Q_{a/b}=\left(\mp\pi/\sqrt{3},-\pi/3\right)$ and 
$\vc Q_c=\left(0,2\pi/3\right)$ touching the Brillouin zone boundary in the 
middle of its edges [see Fig.~\ref{fig:pattern}(g)]. The rotation matrices
have a simple diagonal form, reflecting only the sign changes of the
respective components
\begin{equation}\label{eq:Rmtx4}
\mathcal{R}_{A/B} = 
\mathrm{diag}\,\left(
\pm \ee^{i \vc Q_a\cdot \vc R},
\pm \ee^{i \vc Q_b\cdot \vc R},
\ee^{i \vc Q_c\cdot \vc R}\right) \;.
\end{equation}
The six-sublattice transformation written in Fourier representation has a
full matrix structure
\begin{equation}\label{eq:Rmtx6}
\mathcal{R}_{A/B}=\pm\tfrac13\left(I+M_{A/B}\gamma+M^*_{A/B}\gamma^*\right)
\end{equation}
with the factor
$\gamma = \tfrac13\left( 
\ee^{i \vc Q_1\cdot \vc R} + 
\ee^{i \vc Q_2\cdot \vc R} + 
\ee^{i \vc Q_3\cdot \vc R} \right)$
and the matrices
\begin{equation}
I=\begin{pmatrix}
1 & 1 & 1 \\
1 & 1 & 1 \\
1 & 1 & 1 
\end{pmatrix}
\quad
M_A=\begin{pmatrix}
1 & c^* & c \\
c & 1 & c^* \\
c^* & c & 1
\end{pmatrix}
\quad
M_B=\begin{pmatrix}
c & 1 & c^* \\
1 & c^* & c \\
c^* & c & 1
\end{pmatrix} 
\end{equation}
where $c=\ee^{2\pi i/3}$. The characteristic vectors 
$\vc Q_{1,2}=(-2\pi/3\sqrt{3},\pm 2\pi/3)$ and $\vc Q_3=(4\pi/3\sqrt{3},0)$ 
shown in Fig.~\ref{fig:pattern}(g) again touch the boundary of the Brillouin
zone, now in its corners. The dual transformation takes a general form
$\vc S'_{A\vc R} = \sum_{\vc Q} \ee^{i \vc Q\cdot \vc R}
\mathcal{R}_{A\vc Q} \vc S_{A\vc R}$ (here for sublattice $A$) which
translates into
\begin{equation}\label{eq:dualFour}
\vc S'_{A\vc q}= \sum_{\vc Q} \mathcal{R}_{A\vc Q} \vc S_{A,\vc q-\vc Q} \;,
\end{equation}
\textit{i.e.}, the Fourier components get shifted by the characteristic
vectors. As a side result, the above relation gives the Bragg spots derived
from the Bragg spots of Heisenberg FM/AF ($\vc S_{A,\vc q=0}=\pm \vc S_{B,\vc
q=0}=1$) and presented in Fig.~\ref{fig:pattern}(h).

To study the spin excitations, we employ the spin susceptibility tensor
defined as 
\begin{equation}
\chi_{\alpha\beta}(\vc q,\omega) = 
i\int_0^\infty \langle [S^\alpha_{\vc q}(t),S^\beta_{-\vc q}(0)]\rangle
\,\ee^{i(\omega+i\delta)t}\:\mathrm{d}t \;.
\end{equation}
It is evaluated at the {\SU} points by first decomposing $S_{\vc q}$ 
into the $A$ and $B$-sublattice contributions via
\begin{equation}
\vc S_{\vc q} = \frac1{\sqrt{2}}\,\ee^{i \sqrt{3}q_x/2}
\left(\ee^{iq_y/2} \vc S_{A\vc q} + \ee^{-iq_y/2} \vc S_{B\vc q}\right) \;,
\end{equation}
applying the dual transformation in the Fourier form of Eq.~\ref{eq:dualFour}
to get back to the underlying Heisenberg model, and using the spin
susceptibility for the Heisenberg model obtained within linear-spin-wave (LSW)
approximation. In the case of $\mathcal{T}_4$, this brings simple $\vc
q$-shifts by $\vc Q_a$, $\vc Q_b$, and $\vc Q_c$ for the individual
components. For $\mathcal{T}_6$, the corresponding expressions are somewhat
more involved containing a non-shifted contribution and shifted contributions
combining pairs of the characteristic vectors $\vc Q_1$, $\vc Q_2$, and 
$\vc Q_3$. Without going into details, the presence of both shifted and
non-shifted parts can be easily understood based on Eq.~\ref{eq:Rmtx6}.

%-figure 4-----------------------------------------------------------------
\begin{figure}[tb]
\begin{center}
\includegraphics[scale=1.00]{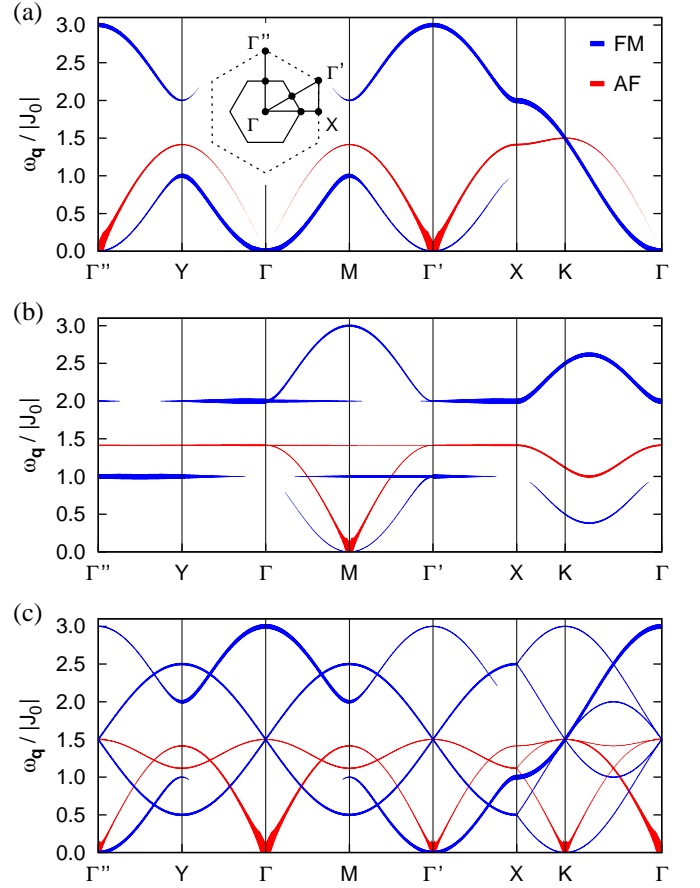}
\caption{(Color online)
(a)~LSW dispersion of the Heisenberg FM (blue) and AF (red) on the honeycomb
lattice. The width of the lines indicates the trace of the spin susceptibility
tensor, $\sum_\alpha \chi''_{\alpha\alpha}(\vc q,\omega)$, calculated in the
LSW approximation.
(b)~The same for the stripy and zigzag state presented in
Fig.~\ref{fig:pattern}(c,d). Energy is scaled by $J_0$ of the hidden
Heisenberg magnet.
(c)~The same for the ``vortex''-like patterns presented in
Fig.~\ref{fig:pattern}(e,f).
}
\label{fig:LSW}
\end{center}
\end{figure}
%--------------------------------------------------------------------------

Presented in Fig.~\ref{fig:LSW} are the traces of the spin susceptibility
tensor of the Heisenberg model and the extended KH model at the two hidden
{\SU} points under consideration. For completeness, we demonstrate both hidden
FM and AF case characterized by quadratic and linearly dispersing Goldstone
modes, respectively. 
The situation is more transparent for the four-sublattice patterns -- stripy
(hidden FM) and zigzag (hidden AF) -- since the spinwave dispersions are just
shifted with the \mbox{$\vc q$=$M$} points replacing the Goldstone points
$\Gamma$ and $\Gamma'$ of the Heisenberg case. In our example, we have chosen
\mbox{$z$-axis} as the ordered moment direction. For the magnons, which are
in fact deviations of the ordered moment in $x$ and $y$ directions, only $\vc
Q_a$ and $\vc Q_b$ shifts are active, selecting four out of the six $M$-points
in total. The remaining two are the Bragg spots reached from $\Gamma$ and
$\Gamma'$ by $\vc Q_c$ shifts active for the ordered $z$ spin component. The
Bragg spots and the Goldstone points are thus complementary in this case.
The spin excitations associated with the six-sublattice patterns are
significantly more complicated. They contain both shifted Goldstone modes 
[in Fig.~\ref{fig:LSW}(c) such a mode appears at \mbox{$\vc q$=$K$-point}
coinciding with $\vc Q_3$] and Goldstone modes at the characteristic momenta
$\vc q=\Gamma$ and $\vc q=\Gamma'$ of the underlying Heisenberg model. In the
latter case just the intensity of the modes has been transferred by the dual
transformation, making \textit{e.g.} the linear Goldstone mode at $\vc
q=\Gamma$ the most intense one in the hidden AF case.

A similar analysis of the spin excitations as presented here for
$\mathcal{T}_4$ and $\mathcal{T}_6$ {\SU} points can be performed for the
remaining {\SU} points. Due to the nature of the relevant transformations, no
other characteristic vectors appear. Therefore, $\vc q=\Gamma$, 
$\vc q=\Gamma'$ and their counterparts shifted by the vectors $\vc Q_{a,b,c}$ 
and $\vc Q_{1,2,3}$ entering the transformations $\mathcal{T}_4$ and
$\mathcal{T}_6$ constitute the entire set of the wavevectors of the Goldstone
modes that can be observed within the extended KH model.

%}}}2

%}}}1

% ==================================================

\section{Application to the real materials}\label{sec:mater}
%{{{1

The aim of this work was to study the basic symmetry properties of the
extended KH model -- a promising spin Hamiltonian for the magnetism of
honeycomb iridates. Below we illustrate how this knowledge, taken together
with the experimental data, helps to locate the plausible windows in otherwise
very large parameter space even for this nearest-neighbor (NN) model. We will
show that, despite having only a single result, the search for full self-dual
transformations $JKDC\leftrightarrow J'K'D'C'$ of the extended KH model
provides us with a surprisingly useful tool in the context of Na$_2$IrO$_3$.
This utility of $\mathcal{T}_1$ emerges due to the recent observation of the
magnetic moment direction \cite{Chu15} which, as we see shortly, imposes an
important constraint on the model parameters. This is because, in general, the
data on magnetic easy axes in a crystal, along with the magnon gaps and torque
magnetometry data, provides a direct information on the symmetry and strength
of the anisotropy terms in spin Hamiltonians, and the case of Na$_2$IrO$_3$ is
of course not any special in this sense. 

To begin with, we recall that Na$_2$IrO$_3$ shows so-called zigzag order,
where the spins on $a$ and $b$ bonds are parallel and form ferromagnetic
chains that run along $X$ direction and couple antiferromagnetically along the
\mbox{$Y$-axis}. This relatively simple collinear magnetic structure has been
first explained \cite{Kim11,Cho12} as due to $2^\mathrm{nd}$-NN $J_2$ and
$3^\mathrm{rd}$-NN $J_3$ Heisenberg couplings (which are often relevant in
compounds with $90^\circ$-bonding geometry -- well known example is quasi-one
dimensional cuprates). This model emphasizes a geometrical frustration which
is realized at large values of $J_{2,3}$ and resolved by the $C_3$ symmetry
breaking zigzag formation. 

However, as argued in the Introduction, more recent data \cite{Gre13,Chu15} 
suggests that the origin of frustrations is largely related to the 
non-Heisenberg-type interactions which are bond-dependent and hence highly 
frustrated even on the level of NN-models. A minimal NN-model of this sort 
is the KH model, which has been shown \cite{Cha13} to host zigzag order 
in its phase diagram indeed. We follow this way of reasoning and 
explore below the extended version of the KH model as the basic NN-model for 
iridates. On the way, we will also see the point where the data may require
the presence of additional terms $J_{2,3}$ too, suggesting that the 
both ``zigzag theories'' above are the part of a full story.

In Ref.~\onlinecite{Cha13}, the available experimental data on Na$_2$IrO$_3$ 
has been fitted using the two-parameter KH model, regarding it as a 
phenomenological spin Hamiltonian with arbitrary parameters. 
For $K=21\:\mathrm{meV}$ and $J=-4\:\mathrm{meV}$, the model was found
consistent with experiments in terms of the type of magnetic ordering,
temperature dependence of static magnetic susceptibility and the low energy
spin-excitation spectrum being compared to powder INS. Later, RIXS
experiments \cite{Gre13,Chu15} confirmed the presence of a high energy branch
of spin excitations, with an even better agreement obtained if the
LSW calculation of Ref.~\onlinecite{Cha13} is replaced by a more
suitable exact diagonalization \cite{Cha15}. 

\begin{figure}[tb]
\begin{center}
\includegraphics[scale=0.96]{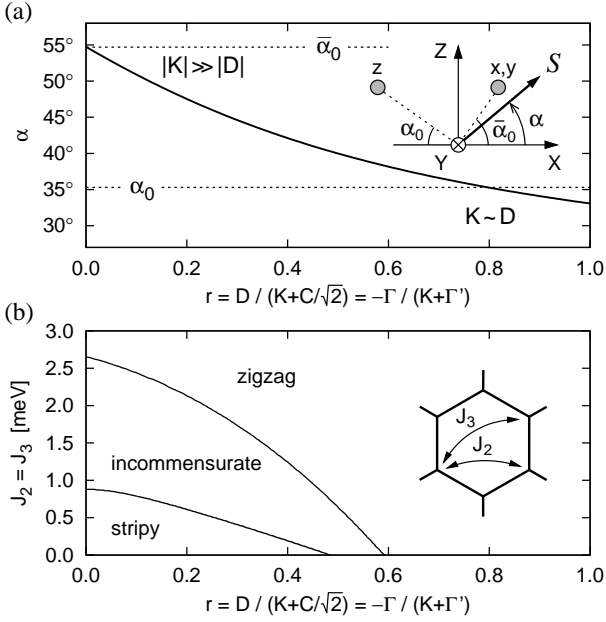}
\caption{
(a) Pseudospin angle $\alpha$ relative to the $XY$-plane (see inset) as 
a function of the parameter $r=D/(K+C/\sqrt{2})=-\Gamma/(K+\Gamma')$. 
Dashed lines show the ``magic'' angle $\alpha_0\simeq 35^\circ$ and its 
complement $\bar{\alpha}_0\simeq 55^\circ$, determined by the \mbox{$z$-axis}
and $xy$-plane, respectively, as sketched in the inset. 
(b)~The phase diagram as a function of long-range couplings $J_2=J_3$ and
anisotropy parameter $r$. Starting with the ``bare'',
\mbox{$\mathcal{T}_1$-derived}
values of $J=5.3\:\mathrm{meV}$, $K=-7.0\:\mathrm{meV}$,
$D=-9.3\:\mathrm{meV}$, and $C=-6.6\:\mathrm{meV}$, we have scaled $D$ and
$C$ simultaneously to vary $r$. To stay within the zigzag phase at the smaller
values of $r\lesssim 0.59$, one needs to have finite $J_{2,3}$ couplings.
Otherwise, the NN-only extended KH model with negative $K<0$ switches to the
incommensurate and ``stripy'' \cite{Cha10,Jia11,Cha13} ground states. The
inset shows the exchange bonds $J_2$ and $J_3$. 
}
\label{fig:mater}
\end{center}
\end{figure}

However, the recent data \cite{Chu15} on the moment direction came about as an
unexpected surprise, challenging at first glance the above coherent 
description of Na$_2$IrO$_3$. The point is that within the original
two-parameter KH model, the zigzag order is characterized by the spins
pointing towards one of the oxygen ions [see Fig.~\ref{fig:schem}(a)]. This
expectation is generic and guaranteed by the
``order-from-disorder'' physics \cite{Tsv95} which typically selects one of
the high-symmetry cubic axes as the easy one, when a spin Hamiltonian
contains the compass-type or Kitaev-type bond-dependent anisotropy
\cite{Kha01,Cha10,Siz14,noteCompass}, independent on parameter values. The
resonant magnetic x-ray diffraction data \cite{Chu15} shows instead that the
magnetic easy axis is in fact far away from any of the Ir-O bond directions:
it is oriented ``nowhere'' slightly below a midpoint between the two, $x$ and
$y$, oxygen ions in Figs.~\ref{fig:schem}(a) and \ref{fig:mater}(a). This is a
clear indication of the significance of the $D$ and $C$ terms in the spin
Hamiltonian \cite{noteEasyaxis}.

To reconcile all the data at hand using now full four-parameter model, we
first notice that the above two easy axis directions -- the one observed in
Na$_2$IrO$_3$ and the one expected from the KH model as used in 
Ref.~\onlinecite{Cha13} -- are roughly related to each other 
simply by a \mbox{$\pi$-rotation} about the $Z$-axis.
This observation gives an immediate hint how to obtain a starting parameter
point when fitting the current data set for Na$_2$IrO$_3$ within the
extended, four-parameter KH model in an appealingly easy way, and resolve
the above apparent problem with the moment direction. 

As discussed in Sec.~\ref{sec:PD}, the $\mathcal{T}_4$-associated zigzag 
phase of the KH model with $K>0$ is related to the zigzag phase connected 
to the {\SU} point $B$ ($\mathcal{T}_1\mathcal{T}_4$) of
Fig.~\ref{fig:PD}(a) via $\mathcal{T}_1$. Remarkably, due to the nature of
$\mathcal{T}_1$ -- a global $\pi$-rotation of the magnetic moments about
$Z$-axis -- all the aforementioned consistent results \cite{Cha13} of the 
two-parameter KH model are fully preserved if we apply Eq.~\ref{eq:fulldual} 
to the parameters $K$ and $J$ of Ref.~\onlinecite{Cha13} given above, 
the only change is the spin easy axis being rotated to the proper 
direction as in experiment. 
The corresponding set of parameters obtained via \eqref{eq:fulldual} is:
$J=5.3\:\mathrm{meV}$, $K=-7.0\:\mathrm{meV}$, 
$D \equiv -\Gamma=-9.3\:\mathrm{meV}$,
$C \equiv \sqrt{2}\Gamma'=-6.6\:\mathrm{meV}$. 
We would like to emphasize that these numbers should not be taken 
literally; rather, they fix the signs of the parameters involved and 
put an upper limit for $D$ and $C$, as we explain below. 

%-figure 5-----------------------------------------------------------------
%--------------------------------------------------------------------------

%-figure 6-----------------------------------------------------------------
\begin{figure}[tb]
\begin{center}
\includegraphics[scale=0.96]{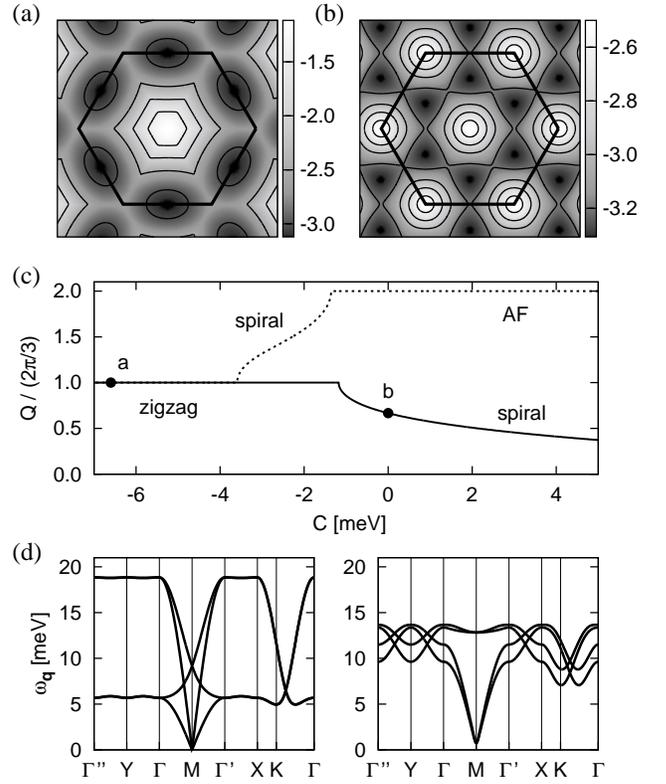}
\caption{
(a)~Map of the $\vc q$-dependent classical energy (per site, in units of meV)
obtained by Luttinger-Tisza method \cite{Lut46,Lit74} for the ``bare'',
\textit{i.e.} \mbox{$\mathcal{T}_1$-derived}
parameters $J=5.3\:\mathrm{meV}$, $K=-7.0\:\mathrm{meV}$, 
$D=-9.3\:\mathrm{meV}$, $C=-6.6\:\mathrm{meV}$ relevant for Na$_2$IrO$_3$.
The hexagon indicates the first Brillouin zone.
(b)~The same for the parameters $K=D=-10\:\mathrm{meV}$, $J=C=0$ 
relevant to Li$_2$IrO$_3$.
(c)~Length of the ordering vector for varying $C$, keeping the other
parameter values unchanged. The dashed (solid) line was calculated using 
the above parameters $JKD$ relevant to Na$_2$IrO$_3$ (Li$_2$IrO$_3$). 
Points $a$ and $b$ show the $C$ values used in panels (a) and (b), respectively.
(d)~LSW dispersions for the parameters used in panel (a) (left) and panel (b)
(right). In the latter case, we have taken $C\simeq -1.2\:\mathrm{meV}$
instead of $C=0\:\mathrm{meV}$ to stay in the zigzag phase at the border to
the spiral phase.
}
\label{fig:mater2}
\end{center}
\end{figure}
%--------------------------------------------------------------------------

For the representative parameters given above, the pseudospin makes a
``magic'' angle of $\alpha_0\simeq 35^\circ$ from the $XY$-plane, as follows
from $\mathcal{T}_1$ construction. This is slightly lower than 
observed \cite{Chu15,noteAlpha}. Now, we inspect how the angle $\alpha$ varies 
as a function of the anisotropy parameters. The result is illustrated 
in Fig.~\ref{fig:mater}(a) and shows that the exact value of $\alpha$ 
heavily influences the ``departure'' from the KH model quantified by 
the parameter $r=D/(K+C/\sqrt{2})=-\Gamma/(K+\Gamma')$. (The corresponding 
Eqs.~\ref{eq:alpha}, \ref{eq:dalpha}, and \ref{eq:alpha2} for the spin 
direction are derived in the Appendix B by minimizing the classical energy). 
The ``magic'' angle $\alpha=\alpha_0$ appears at $r=0.8$ -- as obtained 
for the above parameters. Yet, as observed in Fig.~\ref{fig:mater}(a),
by increasing $\alpha$ for example by $10^\circ$ only, we already find 
$r\simeq 0.3$ and get closer to the $|K|\gg|D|$ regime. 
Therefore, more detailed measurements and fits of the ordered spin direction
are highly desirable to get the actual values of the parameters $D$ and $C$
relative to the Kitaev term $K$. Doing so, it is crucial to take into
account the fact that the pseudospin direction and magnetic moment direction
are not the same in general; while they coincide in the cubic limit, a sizable
trigonal-field splitting might be present in Na$_2$IrO$_3$
\cite{Gre13,noteDelta}. It is thus important to quantify this splitting by
independent measurements. 

At this point, longer-range couplings $J_{2,3}$ become a part of the full spin
model for iridates, for the following reason. As a $\mathcal{T}_1$ partner of
the zigzag phase of Ref.~\onlinecite{Cha13}, the present NN-model with large
$D$ is well in its zigzag ordered state. But this is not so at smaller values
of $D$ ({\it e.g.} for $r\sim 0.5$), which are required to get the spin angles
$\alpha\sim 40^\circ$ or above, see Fig.~\ref{fig:mater}(a). Incorporating
moderate $J_2$ and $J_3$ couplings into the model, we can however stabilize
the zigzag phase, see Fig.~\ref{fig:mater}(b), and hence obtain the ordered
spin angles above the ``magic'' one. 
The values of $J_{2,3}$ of the order of $1-2\:\mathrm{meV}$ are indeed
suggested by ab-initio calculations \cite{Yam14}.
This shows again the key
importance of the experimental data on moment directions for quantifying the
balance between the two zigzag-supporting mechanisms discussed above: 
based on $J_{2,3}$ geometrical frustration, and on frustration driven by the
non-Heisenberg nature of interactions in spin-orbit coupled magnets.
Recent observations \cite{Chu15} of a pronounced spin-space anisotropy
on one hand, and an ``intermediate'' spin direction that requires
finite $J_{2,3}$ values on the other hand, suggest that both
mechanisms are at play in Na$_2$IrO$_3$.

Altogether, the present analysis using the symmetry properties of the model, 
taking into account the recent data on moment direction \cite{Chu15}, 
as well as considering the role of the $J_{2,3}$ couplings 
suggests a plausible window in the parameter space of an effective spin 
model for Na$_2$IrO$_3$: $J_{2,3} < J \sim |C| < |D| < |K|$,
with positive (AF) Heisenberg couplings $J_{2,3}$ and $J$. The leading
anisotropy terms $K<0$ and $D<0$ are both negative, while a smaller term $C$
may in principle take any sign. This parameter window is globally consistent
with experimental observations on Na$_2$IrO$_3$ we are aware of to date, and
may be used as a guide in future analysis, in particular once 
${\vc q}$-resolved spin response becomes available, and the ordered pseudospin 
and magnetic moment directions (they differ in general) are obtained and 
confirmed by independent measurements.

Even though this general result still leaves quite a freedom, it is of great
help by fixing the signs of most relevant couplings and their hierarchy. This
is the main outcome of the present theory in the context of real
materials. Further, we note that the Kitaev coupling $K$ can be deduced
from overall magnetic energy scale, and the spin and magnetic moment
directions should determine the parameter $r$ hence $D$. From a careful
analysis of the zigzag stability condition, magnon gaps and dispersions,
paramagnetic susceptibility data, \textit{etc.}, one should be able to
quantify all the model parameters including $C$, $J$, and $J_{2,3}$.

Considering this result in the context of microscopic theories, we notice
first that the signs of $J>0$ and $K<0$ above are consistent with the original
calculations of these parameters for honeycomb iridates \cite{Jac09,Cha10} as
well as with the later studies \cite{Kat14,Rau14a,Rau14b,Yam14,Siz14}. Next,
we may conclude that a contribution from $t_{2g}-e_g$ hopping that favors
pseudospin interaction with $K>0$ \cite{Kha05} is not significant in iridates;
this is also consistent with the recent calculations \cite{Foy13,Kat14}.
Further, the present symmetry analysis resolves an apparent conflict with the
theoretical $K<0$ \cite{Jac09,Cha10} and the positive $K>0$ that follows from
the best data-fit using the KH model \cite{Cha13}: in fact, the pure KH model
with $K>0$ and the extended one with $K<0$ and sizable $D,C$ terms are
$\mathcal{T}_1$-dual partners (the latter one being physical). 

More surprisingly, relatively large $D(\equiv -\Gamma)$ anisotropy term is
required to ``turn'' the moment direction well away from the pure KH model
position. A positive implication of this observation is that this term makes
it much easier to stabilize the zigzag order (the pure KH model with large
$K<0$ would require large long-range $J_{2,3}$ couplings otherwise). In a
view of the discussion in Sec.~\ref{sec:model}, this suggests a presence of
sizable trigonal field effects in Na$_2$IrO$_3$. Eventually, an unusual --
out of any crystal symmetry axis -- orientation of pseudospins \cite{Chu15}
should originate from a competition among the several anisotropy terms $K$,
$D$, and $C$ of different symmetry and physical origin. 

To conclude our discussion of Na$_2$IrO$_3$: it seems that the extended KH
model, likely further ``extended'' by moderate longer-range couplings, is
indeed a good candidate model for this compound. Even though these extensions
(to be still quantified by future experiments) reduce the chances for ``pure''
Kitaev-model physics in iridates, the model itself is highly interesting due
to its rich internal structure and hidden symmetries that we have uncovered 
in this work. 
 
% Na-213 from PRL2013:               Li-213 from PRL2013:              
%                                   
% K=20.9                             K=15.8
% J=-4.01                            J=-5.3
%                                    
% J'= J+4*K/9 = 5.279                J'= +1.7
% K'= -K/3 = -6.967                  K'= -5.3
% D'= -4*K/9 = -9.289                D'= -7.0
% C'= -2*sqrt(2)*K/9 = -6.568        C'= -5.0
%
% scaled to J,K,Gamma,Gamma':
% 0.41439  -0.54731   0.72714  -0.36489 (Na)
% 0.19009  -0.59263   0.78272  -0.39533 (Li)

Motivated by the above, we further consider the case of Li$_2$IrO$_3$. Since
the data is limited here, the discussion will be brief and suggestive only.
Due to the smaller Curie-Weiss temperature and more ``ferromagnetic'' behavior
of its spin susceptibility \cite{Sin10,Sin12}, this compound was located
closer to {\SU} point of the KH model \cite{Cha13}. Even though the parameters
$K=15.8\:\mathrm{meV}$ and $J=-5.3\:\mathrm{meV}$ given in
Ref.~\onlinecite{Cha13} correspond to the zigzag phase while Li$_2$IrO$_3$
shows a spiral magnetic ordering \cite{Col13}, these parameters can be used to
get a hint of the direction in the parameter space to consider. We therefore
transform the above parameters using \eqref{eq:fulldual} to obtain:
$J=1.7\:\mathrm{meV}$, $K=-5.3\:\mathrm{meV}$, 
$D\equiv -\Gamma=-7.0\:\mathrm{meV}$,
$C\equiv \sqrt{2}\Gamma'=-5.0\:\mathrm{meV}$.
Representing the parameters for Na and Li compounds obtained via
$\mathcal{T}_1$ transformation \eqref{eq:fulldual} in Fig.~\ref{fig:PD}, we
see that both are close to the {\SU} point $B$ ($\mathcal{T}_1\mathcal{T}_4$),
with Li being closer, as expected. To approach the spiral state observed in
Li$_2$IrO$_3$, we first note that, in first approximation, $K\sim D < 0$ in
both cases and that Li compound is characterized by a much smaller $J$. For
simplicity, we set $J=0\:\mathrm{meV}$, assume $K=D=-10\:\mathrm{meV}$ to
roughly preserve the overall energy scale, and reduce the parameter $C$
associated with the trigonal distortion, which is expected to be much smaller
in Li$_2$IrO$_3$ with the bond angles being closer to $90^\circ$. The
Luttinger-Tisza \cite{Lut46,Lit74} maps of the classical energy for the Na and
Li case presented in Fig.~\ref{fig:mater2}(a,b) confirm the zigzag and
incommensurate magnetic ordering, respectively. For $C=0$ the incommensurate
ordering wavevector is obtained as $\vc Q\simeq\frac23 \vc Q_c$ [see
Fig.~\ref{fig:mater2}(b,c)], this would predict a magnetic Bragg peak in
powder neutron diffraction experiments at a $|\vc Q|$-value that could be
consistent with experiments on powder Li$_2$IrO$_3$ \cite{Col13}.
Finally, Fig.~\ref{fig:mater2}(d) compares the spin excitations obtained using
LSW approximation. In the case of Na$_2$IrO$_3$, the dispersion
is identical to that presented in Fig.~3 of Ref.~\onlinecite{Cha13},
possessing a low and high-energy branches. As the parameter $J$ is reduced,
these two branches gradually merge, leading to a steeper dispersion 
compared to Na$_2$IrO$_3$, which might be consistent with powder inelastic 
neutron scattering experiments on powder Li$_2$IrO$_3$ \cite{Col13}. The 
predicted dispersion is illustrated in Fig.~\ref{fig:mater2}(d) for a
point on the boundary between the zigzag and the spiral phase. A further 
minor reduction of $C$ to enter the spiral phase and get the proper 
ordering vector should not affect this result dramatically, apart from 
the changes at low energies forming an ``hour-glass'' shape characteristic 
to spiral magnets (see, \textit{e.g.}, Ref.~\onlinecite{Kim14c}).

%}}}1

% ==================================================

\section{Conclusions}

To summarize, we have analyzed non-trivial symmetries of the extended
Kitaev-Heisenberg model on the honeycomb lattice. As a main result, we have
identified the complete set of points in the parameter space where this
bond-anisotropic model can be transformed to a simple Heisenberg model and is
therefore characterized by hidden {\SU} symmetry. Such a dual transformation
can be performed using a particular choice of sublattice rotations of the
spins, specific for each of the {\SU} points.
The sublattice structure of the transformations creates a number of ordering
patterns which together with the location of the hidden {\SU} points in the
parameter space give a good overview of the global phase diagram of the model.
In terms of the spin excitations, the hidden {\SU} symmetry manifests itself
by the presence of Goldstone modes inherited from the {\SU} symmetric
Heisenberg FM/AF on the honeycomb lattice. Their characteristic vectors and
even the full spin excitation spectra are easily obtained by an explicit
transformation of the FM/AF case. 

One of the special transformations linked to the hidden {\SU} points reveals
at the same time an exact duality between the Kitaev and Kekul\'{e}-Kitaev
models; this result should be useful in theoretical studies of these and
related models.

We emphasize that, adopting the extended KH model, all the above results are
necessary consequences of its symmetry which is in turn dictated by the
underlying $C_3$ symmetry of the lattice.

Having the results of the general symmetry analysis at hand, we were able to
find the region of the parameter space that is consistent with the observed
properties of the honeycomb lattice iridates Na$_2$IrO$_3$ and Li$_2$IrO$_3$.
Further, a relation between the ordered moment direction and the model
parameters is derived, which may help to quantify these parameters from future
experiments.

Finally, our method to systematically explore the hidden symmetries is general
and can be applied to other bond-anisotropic models as well. In the context of
the iridate materials, the symmetry analysis of the extended KH model on
hyper-honeycomb and harmonic honeycomb lattices is of a great interest.

% ==================================================

\acknowledgments

We would like to thank B.J.~Kim for sharing with us the experimental 
data \cite{Chu15} which motivated this study, R.~Coldea, 
G.~Jackeli, I.~Kimchi, N.B.~Perkins, S.~Trebst, and S.E.~Sebastian for 
helpful discussions and comments.
JC acknowledges support by ERDF under project CEITEC (CZ.1.05/1.1.00/02.0068),
EC 7$^{\rm{th}}$ Framework Programme (286154/SYLICA), and Czech Science
Foundation (GA\v{C}R) under project no. 15-14523Y.

\appendix

\section{XYZ form of the Hamiltonian}
%{{{1

The Hamiltonian expressed in terms of the spin components $S^X$, $S^Y$, and
$S^Z$, corresponding to the $XYZ$ reference frame in Fig.~\ref{fig:schem}(a),
takes the form
\begin{align}\label{eq:Hglob}
& \mathcal{H}_{\langle ij\rangle\, \parallel\, \gamma} =
J_{XY} (S_i^X S_j^X + S_i^Y S_j^Y) + J_Z S_i^Z S_j^Z \notag \\
& + A\, [ c_\gamma (S_i^X S_j^X\!-\!S_i^Y S_j^Y) -
          s_\gamma (S_i^X S_j^Y\!+\!S_i^Y S_j^X) ] \notag \\
 & - B\,\sqrt{2}\,[c_\gamma (S_i^X S_j^Z\!+\!S_i^Z S_j^X) +
                   s_\gamma (S_i^Y S_j^Z\!+\!S_i^Z S_j^Y) ] \;.
\end{align}
Here the $C_3$ symmetry of the model is embodied in the factors 
$c_\gamma\equiv\cos\phi_\gamma$ and $s_\gamma\equiv\sin\phi_\gamma$,
where the angles $\phi_\gamma$ are determined by the bond directions:
$\phi_\gamma=0,\frac{2\pi}3,\frac{4\pi}3$ for the $c$, $a$, and $b$ bonds,
respectively. In terms of the original parameters $JK\Gamma\Gamma'$,
the exchange constants entering \eqref{eq:Hglob} read as
\begin{align}
&A      = \tfrac13 K+ \tfrac23 (\Gamma-\Gamma')\;, \\
&B      = \tfrac13 K- \tfrac13 (\Gamma-\Gamma')\;, \\
&J_{XY} = J+B-\Gamma'\;, \\
&J_Z    = J+A+2\Gamma'\;.
\end{align}

Note that it is the $A$ and $B$ terms which bring about the
bond-directionality of the interactions, and hence they naturally support
$C_3$-symmetry breaking orderings such as zigzag in the present model.
Physically, these terms arise from the exchange processes that involve the
in-plane components of orbital momentum $L_X$ and $L_Y$ which ``know'' the
bond directions, like the orbitals do in the Kugel-Khomskii models. 

It is also noticed that the $A$ and $B$ terms change the $Z$-component of
total angular momentum by $\pm 2$ and $\pm 1$, correspondingly. This is because 
the $t_{2g}$-orbital angular momentum $L$ is not a conserved quantity in a
crystal, and this commonly shows up in effective spin Hamiltonians due to the
spin-orbit coupling. 

A strong trigonal field splits $t_{2g}$-level such that the lowest 
Kramers doublet (pseudospin) wavefunctions 
$|\tilde\uparrow\rangle,|\tilde\downarrow\rangle$ become simple 
products of $L_Z=\pm 1$ and spin $|\!\!\downarrow\rangle$, 
$|\!\!\uparrow\rangle$ states, correspondingly; \textit{i.e.}, there 
will be a one-to-one correspondence between the real spin and pseudospin 
directions. Since the total spin is conserved during the hoppings, 
pseudospin is then conserved, too. Thus, the spin non-conserving terms 
$A$ and $B$ must vanish in this limit, which implies $K\rightarrow 0$ 
and $\Gamma\rightarrow\Gamma'$ simultaneously. Physically, 
a strong compression along trigonal axis dictates that this axis becomes the
``easy'' (or ``hard'') one for moments. Since this limit is not realized in
iridates, we will not use the $XYZ$-form of Hamiltonian in this paper;
however, it might be useful for pseudospin one-half Co$^{4+}$, Rh$^{4+}$, 
and Ru$^{3+}$ compounds where the spin-orbit and crystal field effects may
strongly compete. 

%}}}1

\section{Analysis of the classical energy and moment direction}
%{{{1

In this appendix we show the expressions used in the classical energy
analysis. We first give the Hamiltonian in its momentum space form utilized
within the Luttinger-Tisza method. By minimizing the classical energy in the
zigzag phase we then find the ordered moment direction.

Transforming the spin operators via 
$\vc S_{A\vc R}=\sum_{\vc q} \ee^{i\vc q\cdot\vc R} \vc S_{A\vc q}$ and
similarly for the $B$-sublattice, we cast the Hamiltonian to the form
\begin{equation}\label{eq:Hamq}
\mathcal{H} = \sum_{\vc q} 
\Psi_{\vc q}^\dagger\, H_{\vc q} \Psi_{\vc q}^{\phantom{\dagger}}
\quad\text{with}\quad
\Psi_{\vc q}=
\begin{pmatrix}
\vc S_{A\vc q} \\
\vc S_{B\vc q}
\end{pmatrix} \;,
\end{equation}
where the $\vc q$-vectors cover the first Brillouin zone of the triangular
lattice of $\vc R$. The simplest expressions for the $6\times6$ matrices 
$H_{\vc q}$ of the momentum-space Hamiltonian \eqref{eq:Hamq} are obtained 
using the cubic axes $x$, $y$, $z$. Complementing the interactions in 
Eq.~\ref{eq:Hcubic} by long-range
$J_2$ and $J_3$, we arrive at
\begin{equation}
H_{\vc q} = 
N_\mathrm{site} 
\begin{pmatrix}
F_{\vc q} & G_{\vc q} \\
G^\dagger_{\vc q} & F_{\vc q} \rule{0pt}{12pt}
\end{pmatrix}
\quad\text{with}\quad
F_{\vc q} = \tfrac12 J_2 
\begin{pmatrix}
1 & 0 & 0 \\
0 & 1 & 0 \\
0 & 0 & 1 
\end{pmatrix} \,\eta_{2\vc q}
\end{equation}
and
\begin{multline}
G_{\vc q} = \tfrac14 
(J_1\eta_{1\vc q} + J_3\eta_{3\vc q}) \begin{pmatrix}
1 & 0 & 0 \\
0 & 1 & 0 \\
0 & 0 & 1 
\end{pmatrix}
+ \tfrac14 K \begin{pmatrix}
e_1 & 0 & 0 \\
0 & e_2 & 0 \\
0 & 0 & 1 
\end{pmatrix} \\
+ \tfrac14 \Gamma \begin{pmatrix}
0   & 1   & e_2 \\
1   & 0   & e_1 \\
e_2 & e_1 & 0 
\end{pmatrix}
+ \tfrac14 \Gamma' \begin{pmatrix}
0       & e_1+e_2 & e_1+1 \\
e_1+e_2 & 0       & e_2+1 \\
e_1+1   & e_2+1   & 0 
\end{pmatrix}\;.
\end{multline}
Here the momentum-dependent factors read as
\begin{align}
e_{1,2} &=\ee^{-i\frac12(\pm\sqrt3q_x+3q_y)} \;, \\
\eta_{1\vc q} &= 
 1+2\cos\tfrac{\sqrt{3}q_x}2\,\ee^{-i\frac32 q_y} \;, \\
\eta_{2\vc q} &= 
 \cos\sqrt{3}q_x+2\cos\tfrac{\sqrt{3}q_x}2\,\cos\tfrac{3q_y}2 \;, \\
\eta_{3\vc q} &= 
 \ee^{-i 3q_y}+2\cos\sqrt{3} q_x \;.
\end{align}

In the Luttinger-Tisza method \cite{Lut46,Lit74}, the matrices 
$H_{\vc q}/2N_\mathrm{site}$ for ${\vc q}$ running through the Brillouin 
zone are diagonalized. The \mbox{${\vc q}$-vector} and the eigenvector 
corresponding to the minimum eigenvalue then determine the ordering resulting
on a classical level and the minimum eigenvalue itself gives the classical
energy per site. This approach relaxes the spin-length constraint which should
be checked afterward.

Next, we evaluate the classical energy for the zigzag state with the ordering
vector $\vc Q=\vc Q_c=(0,2\pi/3)$ and an arbitrary ordered moment direction 
given by a unit vector $\vc u$. The corresponding zigzag pattern is captured
by $\Psi_{\vc Q}=(+\frac12\vc u,-\frac12 \vc u)^T$.
Using \eqref{eq:Hamq}, we get for the classical energy per site:
\begin{equation}
E_\mathrm{class}=\tfrac18(J_1-K-2J_2-3J_3)+ \tfrac18 \vc u^T M \vc u
\end{equation}
with the matrix
\begin{equation}
M=
\begin{pmatrix}
2K               & -\Gamma+2\Gamma' & \Gamma \\
-\Gamma+2\Gamma' & 2K               & \Gamma \\
\Gamma           & \Gamma           & 0        
\end{pmatrix}
\end{equation}
or equivalently
\begin{equation}
M=
\begin{pmatrix}
2K          & D+\sqrt{2}C & -D \\
D+\sqrt{2}C & 2K          & -D \\
-D          & -D          & 0        
\end{pmatrix} \;.
\end{equation}

The ordered moment direction can now be obtained as the eigenvector of $M$
corresponding to its lowest eigenvalue. However, as will be clear in a moment, 
it is more convenient to switch to the reference frame which coincides with 
the local $\tilde{x}$, $\tilde{y}$, $\tilde{z}$ axes for 
$c$-bonds [see Fig.~\ref{fig:schem}(b)]. 
The matrix $M$ is then transformed to
\begin{equation}
\tilde{M}=
\begin{pmatrix}
2K\!-\!D\!-\!\sqrt{2}C & 0                      & 0          \\
0                      & 2K\!+\!D\!+\!\sqrt{2}C & -\!\sqrt{2}D \\
0                      & -\!\sqrt{2}D           & 0        
\end{pmatrix}
\end{equation}
which can be readily diagonalized and the angle $\alpha$ of the ordered 
pseudospin to the \mbox{$XY$-plane} can be found. 

As discussed in the main text, if we rotate the spins by $180^\circ$ around 
the global \mbox{$Z$-axis}, the observed moment would come close to the 
\mbox{$\tilde{z}$-axis}. 
Since the latter is an attractive point for the two-parameter KH 
model \cite{Cha10}, we guess that this rotation will transform the actual 
$J,K,D,C$ Hamiltonian ($K<0$, large $D$) for Na$_2$IrO$_3$ into 
an effective $J',K',D',C'$ one, with $K>0$ and small only $D'$ and $C'$ 
values, \textit{i.e.} into a nearly two-parameter KH model (which guarantees 
that the corresponding effective easy axis is close to $\tilde{z}$). We 
therefore first apply $\mathcal{T}_1$ transformation via Eq.~\ref{eq:fulldual}, 
calculate the moment direction for effective $J',K',D',C'$, and later 
make use of the expected smallness of the transformed $D'$. 
The first two steps yield an analytical expression for the angle $\alpha$: 
\begin{equation}
\alpha=\alpha_0+\frac12\arctan\left(
\frac{2\sqrt{2}D'}{2K'+D'+\sqrt{2}C'} \right)
\end{equation}
with the first contribution being the ``magic'' angle 
$\alpha_0=\arcsin(1/\sqrt{3})\simeq 35.3^\circ$ of 
the \mbox{$\tilde{z}$-axis} to the \mbox{$XY$-plane} and the second 
contribution supposed to be small. Now, we return to the original spin
axes by applying the $\mathcal{T}_1$ transformation again. 
This does not alter the angle $\alpha$ but rotates the moment into its
physical position: below a midpoint between two oxygen ions \cite{Chu15}.
In terms of the original parameters we have
\begin{equation}\label{eq:alpha}
\alpha=\alpha_0+\frac12\arctan\left(
2\sqrt{2}\,\frac{4K-5D+2\sqrt{2}C}{14K+23D+7\sqrt{2}C} \right) \;.
\end{equation}
For $D=C=0$, this equation gives the moment direction towards a
midpoint of two oxygens, as expected for negative $K$ values of the
Kitaev coupling \cite{Chu15} on a classical level (but it will turn to
either $x$- or $y$-oxygen direction once the order-by-disorder mechanism is
switched on \cite{Cha10,Siz14}). The moment moves down from this position once
the model is extended by $D$ and $C$ terms of a proper sign. At the
parameter set given in the main text, the moment takes the ``magic''
angle. By expanding the arctangent near this point, we arrive at the 
following formula for the deviation from $\alpha_0$:
\begin{equation}\label{eq:dalpha}
\delta\alpha=\frac12\arctan\left(
\frac{4\sqrt{2}}7 \frac{1-\frac54 r}{1+\frac{23}{14} r}\right)
\approx\frac{2\sqrt{2}}7 \frac{1-\frac54 r}{1+\frac{23}{14} r}
\end{equation}
with the single parameter
\begin{equation}
r=\frac{D}{K+\frac1{\sqrt 2} C}=-\frac{\Gamma}{K+\Gamma'} \;.
\end{equation}
This parameter quantifies the ``departure'' from the KH model and can be
measured by resonant x-ray \cite{Chu15} or neutron diffraction experiments; 
as mentioned in the main text, care has to be taken in the fits by considering 
the crystal-field effects on pseudospin wavefunctions. 

In terms of the parameter $r$, Eq.~\ref{eq:alpha} can be rewritten as 
\begin{equation}\label{eq:alpha2}
\tan 2\alpha=4\sqrt{2}\,\frac{1+r}{7r-2} \;. 
\end{equation} 
Note, that this and the previous equations for $\alpha$
hold at finite $J_{2,3}$ Heisenberg corrections as well, since 
the easy axis is determined solely by the anisotropy terms. 

%}}}1

\vfil


\begin{thebibliography}{99}
\bibitem{noteDelta}
A relative strength of the trigonal field splitting $\Delta/\lambda$ in
Na$_2$IrO$_3$ can be deduced from a splitting 
$\Delta_{BC}\sim 0.1\:\mathrm{eV}$ \cite{Gre13a} of the excited $J=3/2$ 
quartet, using a relation 
$\Delta_{BC}/\lambda=\frac{1}{\sqrt{2}}\tan^{-1}\theta -1$, where the angle 
$\theta$ is given by $\tan 2\theta=2\sqrt{2}/(1+\frac{2\Delta}{\lambda})$. 
This gives $\Delta/\lambda \simeq 3/8$. With this ratio, we 
obtain the pseudospin $g$ factors (see Ref.~\onlinecite{Abr70} for details) 
$g_c\simeq -2.6$ and $g_{ab}\simeq 1.7$ for out-of-plane and in-plane 
directions, correspondingly, in fair agreement with the observed values 
$|g_c|\simeq 2.7$ and $|g_{ab}|\simeq 1.9$ \cite{Sin10}.
\bibitem{noteKpts}
Unfortunately, the requirement of the Hamiltonian being represented by
non-singular matrices prevents us to find all the hidden Kitaev points by
using the transformations $K_0\leftrightarrow JKDC$. We can only reveal a 
part of them by using $J_0K_0\leftrightarrow JKDC$ and setting $J_0=0$.
\bibitem{noteAlpha} 
Ref.~\onlinecite{Chu15} has estimated the angle $\alpha\simeq 44^\circ$ based 
on the fits that did not account for a trigonal field splitting. Positive 
(negative) $\Delta$ values may decrease (increase) the actual value of 
$\alpha$ from this estimate.
\end{thebibliography}
\end{document}